\documentclass[rmp,aps,superscriptaddress,a4paper,twocolumn,showpacs]{revtex4} 
\usepackage{graphics,epsfig,amsmath,natbib}
\usepackage{setspace}
\newcommand{\be}{\begin{equation}}
\newcommand{\ee}{\end{equation}}
\newcommand{\bea}{\begin{eqnarray}}
\newcommand{\eea}{\end{eqnarray}}
\newcommand{\ba}{\begin{eqnarray*}}
\newcommand{\ea}{\end{eqnarray*}}
\newcommand{\dagga}{{\phantom{\dagger}}}

\newcommand{\bQ}{\mathbf{Q}}

\newcommand{\bk}{\mathbf{k}}

\begin{document}
\title{Modeling the Unconventional Superconducting Properties of Expanded A$_3$C$_{60}$ Fullerides}

\author{Massimo Capone}
\affiliation{SMC, CNR-INFM and Dipartimento di Fisica, Universit\`a ``La Sapienza'', P.le Aldo Moro 2,\\ 
I-00185, Roma, Italy}
\affiliation{ISC-CNR, Via dei Taurini 19, I-00185 Roma, Italy}
\author{Michele Fabrizio}
\affiliation{International School for Advanced Studies (SISSA), and CNR/INFM DEMOCRITOS\\
National Simulation Center, Via Beirut 2-4, I-34014 Trieste,Italy}
\affiliation{The Abdus Salam International Center for Theoretical Physics 
(ICTP), P.O.Box 586,\\ I-34014 Trieste, Italy}
\author{Claudio Castellani}
\affiliation{SMC, CNR-INFM and Dipartimento di Fisica, Universit\`a ``La Sapienza'', P.le Aldo Moro 2,\\ 
I-00185, Roma, Italy}
\author{Erio Tosatti}
\affiliation{International School for Advanced Studies (SISSA), and CNR/INFM DEMOCRITOS\\
National Simulation Center, Via Beirut 2-4, I-34014 Trieste,Italy}
\affiliation{The Abdus Salam International Center for Theoretical Physics 
(ICTP), P.O.Box 586,\\ I-34014 Trieste, Italy}

\date{\today}
\begin{abstract}

The trivalent alkali fullerides solids of generic composition A$_3$C$_{60}$,  
where C$_{60}$ is the fullerene molecule and A = K, Rb, and Cs, are a well established 
family of molecular superconductors. The superconductive electron pairing is
of regular $s$-wave symmetry and is accounted for by conventional coupling 
of electrons to phonons, in particular by well understood Jahn Teller intramolecular 
C$_{60}$ vibrations. A source of renewed interest in these systems are alarming 
indications of strong electron-electron repulsion phenomena, which emerged especially 
in compounds where the C$_{60}$-C$_{60}$ distance is expanded, by either a large cation 
size or by other chemical or physical means. Several examples are now known where 
this kind of expansion, while leading to a high superconducting temperature at first, 
gradually or suddenly causes a decline of superconductivity and its eventual
disappearance in favor of a Mott insulating state. This kind of insulating state 
is the hallmark of strong electron correlations in cuprate and organic superconductors, 
and its appearance suggests that fullerides might also be members at large of that family.\\ 

Our approach to the fullerides is theoretical, and based on the solution of a Hubbard 
type model, where electrons hop between molecular sites. We take advantage of the fact 
that, unlike models for the strongly correlated cuprates, still under debate, in a Hubbard 
model of fullerides all the important electron correlations occur within the molecular site, 
efficiently soluble in the Dynamical Mean Field Theory (DMFT) approximation. 
DMFT solutions confirm that superconductivity in this model fulleride, although of $s$-wave 
symmetry rather than $d$-wave, shares many of the properties that are 
characteristic of high $T_c$ cuprates. The calculations are heavy; and while our 
working model is several years old, the new results we present in this Colloquium 
pertain to the most interesting case of three electrons per C$_{60}$ molecule, 
appropriate to $A_3$C$_{60}$, and have only been possible recently thanks to 
a stronger computational effort.\\

We have calculated the zero temperature phase diagram as a function of the 
ratio of intra-molecular repulsion parameter $U$ over the electron bandwidth $W$, the
increase of $U/W$ representing the main effect of lattice expansion. The phase diagram is 
close to that of actual materials, with a dome shaped superconducting order parameter
region preceding the Mott transition for increasing cell volume. Unconventional properties 
of expanded fulleride superconductors predicted by this model include: 
(i) an energy pseudogap in the normal phase; (ii) a gain of electron kinetic energy 
and of conducting Drude weight at the onset of superconductivity, as in high $T_c$ 
cuprates; (iii) a spin susceptibility and a specific heat behavior that is not 
drastically different from a regular phonon superconductor, despite 
strong correlations; (iv) the emergence of more than one energy scale governing the 
renormalized single particle dispersion, electronic entropy and the specific heat jump. 
These predictions, which if confirmed should establish fullerides, especially 
the expanded ones, as members of the wider family of strongly correlated superconductors, 
are discussed in the light of existing and foreseeable experiments.

\end{abstract}
\pacs{71.30+h, 71.10.Pm, 71.10.Fd}
\maketitle
\tableofcontents

\section{introduction}
\label{introduction}

Discovered by Kamerlingh Onnes nearly a century ago~\cite{onnes}, and first explained 
microscopically back in 1957 in terms of electron pairing by Bardeen,Cooper and Schrieffer (BCS)~\cite{BCS,BCS-1}, 
superconductivity is still surprisingly {\it a' la page}. On one 
hand, superconductivity is being constantly discovered in an ever increasing variety of 
solid state compounds. On the other hand, it appears more and more difficult to use 
basically the same standard theory, essentially BCS and its extensions~\citep[e.g.][chapters~10~and~11]{BCSreview}
to account for all of them. In this standard, conventional theory, superconductivity
arises from the condensation of electron pairs, the two electrons usually bound in 
a pair state of $s$-wave symmetry and held together by exchange of lattice
phonons. The Coulomb repulsion between the two electrons opposes pair formation, but
it does not suppress superconductivity because screening makes it weak enough.

The surprisingly favorable effect of repulsive electron correlations on superconductivity 
found in some systems, particularly in high-$T_c$ superconducting cuprates, where 
electron-electron repulsion is dominant, remains a standing puzzle. An immense amount 
of experimental and theoretical work has accumulated over the last two decades in the 
attempt to understand these phenomena, see e.g. the review by \textcite{hitc}.  Actually, cuprates are but the most 
spectacular members of a wider class of strongly correlated superconductors including 
heavy fermion and organic molecular compounds~\cite{hitc}, systems for which 
there is no really comprehensive theory either. Among other factors, theoretical efforts 
have been hampered by the general {\it inter-site} nature of electron interactions 
and correlations in many of these systems, a fact that poses large technical 
difficulties. In this light, identifying a superconductor family where correlations are 
at the same time strong, simple, and {\it on-site} is welcome.  

A more crucial element is one of perspective. It has been a widespread prejudice to 
distinguish between superconductors where (as in BCS theory) pairing of electrons 
takes place in the $s$-wave channel and is mediated by phonons, from those where 
the mechanism may be electronic and not phononic, and where pairing instead takes
place in the $d$-wave channel. Whereas it is widely held that strong repulsive 
correlations are essential to superconductivity in the latter~\cite{PWA}, they 
are not considered crucial in the former. The conventional BCS scenario and its 
extensions, namely the Migdal-Eliashberg theory~\cite{Migdal,Eliashberg,BCSreview} -- a controlled approximation valid 
when the typical phonon frequency is much smaller than the Fermi-energy -- are more 
or less automatically accepted, and used to account for the superconducting properties. 
In this theory, electron-electron repulsion merely renormalizes the electron-phonon 
parameters, lowering the critical temperature $T_c$ rather than enhancing it.

The trivalent alkali fullerides superconductors, excellently reviewed, e.g., by 
~\textcite{Gunnarsson-review,Gunnarsson-book} and \textcite{Ramirez}, are among the systems where 
this conventional logic seemingly applies. Fullerides  are solid state compounds of generic 
composition A$_3$C$_{60}$, where C$_{60}$ is the fullerene molecule~\cite{Gunnarsson-book} 
and A = K, Rb, and Cs are alkali cations. The three alkalis donate a total of 
$n$=3 electrons to each fullerene, half filling its threefold degenerate $t_{1u}$ 
molecular level. Electron hopping between first neighboring fullerenes gives
rise to a metal, where conduction is restricted to the three narrow $t_{1u}$-derived 
bands, with a total energy bandwidth of no more than 0.6 eV~\cite{satpathy,erwin}.
Metallic fullerides are generally superconducting, with critical temperatures 
$T_c$ reaching $\sim$ 40~K, depending on various factors. An empirically 
important factor appears to be the cell volume. When the 
fulleride lattice is chemically expanded, by either increasing cation size or 
by insertion of neutral molecules, or else physically expanded by removing pressure,
$T_c$ undergoes a definite and systematic change. It rises initially with 
a good correlation with the C$_{60}$-C$_{60}$ distance~\cite{fischer,Gunnarsson-review}. 
Further expansion however causes $T_c$ to drop, ending eventually, through a first order 
transition, in an insulating state, as we shall discuss later. 

A wealth of evidence indicates that superconducting pairing in fullerides
is phononic and that the relevant phonons are the stiff intra-molecular $H_g$ vibrations 
of the C$_{60}$ molecule, Jahn-Teller coupled to the $t_{1u}$ conduction 
electrons~\cite{Gunnarsson-review}. Further support to the apparent BCS 
nature of superconductivity in fullerides comes from specific-heat jumps 
that scale linearly with $T_c$ in agreement with BCS theory~\cite{Ramirez,Ramirez-1,Ramirez-2}, 
as well as a regular (i.e., not exceedingly high) normal phase magnetic susceptibility.~\cite{Petit-Robert,Ramirez-1} 
Superconducting energy gaps are less clearly defined,~\cite{Gunnarsson-review} 
the gap ratio $2\Delta/T_c \simeq 3.4-4.2$ in K$_3$C$_{60}$ and 
Rb$_3$C$_{60}$~\cite{Gunnarsson-review, Ramirez}, but not far from the BCS value 
of 3.53. These elements suggest viewing the A$_3$C$_{60}$ compounds as weakly 
correlated Fermi-liquid conductors~\cite{Ramirez}, though with unusually narrow 
electron bands, with a large effective mass roughly three free electron masses~\cite{Petit-Robert}. 
Even the observed decrease of $T_c$ under applied pressure 
in K$_3$C$_{60}$ and Rb$_3$C$_{60}$ is in qualitative agreement with 
an increasing bandwidth and decreasing density of states at the Fermi level, 
which further supports a standard BCS picture.

These reassuring, conventional looking facts are however contrasted by
a number of conflicting elements that are strong enough to cast serious 
doubts on the general applicability of the BCS scenario to superconductors
in this family. These elements are especially apparent in 
the more expanded fullerides including (NH$_3$)$_x$NaK$_2$C$_{60}$~\cite{Ricco_intercalato} 
and Li$_3$C$_{60}$~\cite{Durand2003}, and 
in the alloys Cs$_{3-x}$K$_x$C$_{60}$ and Cs$_{3-x}$Rb$_x$C$_{60}$~\cite{Dahlke}.
For these expanded compounds $T_c$ decreases upon expansion, contrary to BCS 
theory. The electron density of states extracted by NMR Knight shift 
is at the same time an increasing function of lattice parameter, 
smoothly connecting with that of the unalloyed compounds under pressure~\cite{Dahlke}.
Within BCS theory, that increase should lead to a rise of $T_c$ and not to a 
drop as observed.  The same unconventional behavior is observed in Cs$_3$C$_{60}$, the 
fulleride compound with the highest $T_c\sim 40$~K attained under pressure~\cite{Palstra}. 
A novel A15 superconducting phase of Cs$_3$C$_{60}$ with expanded structure 
has very recently been synthesized~\cite{Ganin} corresponding 
to a body-centered cubic arrangement of fullerenes. Superconductivity 
emerges under pressure through a first order non-structural transition at 4~Kbar.
The critical temperature $T_c$ first increases with pressure, reaching a dome-shaped 
maximum of 38~K around 7~Kbar, above which $T_c$ drops. Since no structural 
changes are observed under pressure, the appearance of superconductivity as well 
as the dome-shaped $T_c$ vs. pressure behavior must be ascribed solely to the volume 
contraction~\cite{Ganin}. This nonmonotonic behavior of $T_c$ with pressure
finds no apparent explanation within the conventional theory. 

The basic and striking anomaly of expanded fullerides occurs in the 
compounds with the largest inter-molecule distances. In these materials a relatively modest additional lattice expansion 
(and minor change of symmetry due to intercalated ammonia) is enough to dramatically 
turn them from metallic and superconducting to antiferromagnetic and insulating~\cite{Iwasa,Durand2003}.\footnote
{We expect that below the superconducting pressure of 4 kbar, 
the new compound A15 Cs$_3$C$_{60}$~\cite{Ganin}, yet to be characterized in this respect, 
should also be an insulating antiferromagnet.}
With temperature, antiferromagnetic order in the ammoniated compound NH$_3$K$_3$C$_{60}$
changes to paramagnetic disorder at a N\'eel temperature slightly above $\sim$ 40~K~\cite{Prassides1999}.
Even above the N\'eel temperature, the microwave conductivity in NH$_3$K$_3$C$_{60}$ remains 
several orders of magnitude below that of K$_3$C$_{60}$~\cite{Kitano}, 
testifying the Mott insulator (correlation driven) nature of the insulating phase. 
Electrons in a lattice give rise to a Mott insulating state when electron electron 
repulsion stops their free propagation and the lattice appears as a collection of molecular ions. 
Correlations lead to an energy gap in their 
spectrum, even if their number density per cell is odd, instead of even as in
regular band insulators~\cite{Mott}.
Proximity of a Mott insulator phase in fullerides had long been advocated
in different contexts~\cite{Kivelson,Kivelson-1,Baskaran,lof} but was not taken seriously 
by the community prior to this data. Superconductivity next to a Mott insulating phase 
as a function of doping or volume change is the hallmark of strong correlations 
in high temperature superconducting cuprates  and organics. 
One is thus naturally led to inquire whether superconductivity in expanded 
fullerides might, despite the obvious differences, and despite the phononic mechanism,
be somehow related to strong correlations. Our contention is that it is indeed 
closely related, as outlined in the following.
\begin{figure}
\includegraphics[width=8.5cm]{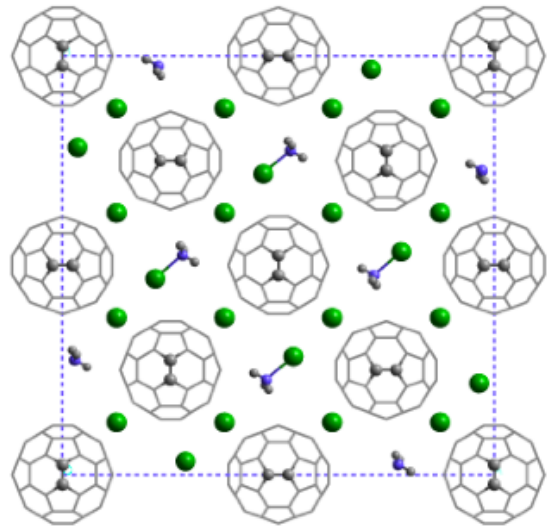}
\caption{\label{structure} Schematic representation of a planar projection of the crystal 
structure of NH$_3$K$_3$C$_{60}$. Green dots are potassium atoms, blue dots surrounded by 
three grey dots are NH$_3$ molecules, and C$_{60}$ molecules are shown according 
to their actual spatial orientation. (Courtesy of Kosmas Prassides).}
\end{figure} 

\section{Electron Interactions in Fullerides}
The electrons donated by the alkalis to the C$_{60}$ molecule enter the threefold 
degenerate $t_{1u}$ former lowest unoccupied molecular orbital (LUMO). In a
degenerate molecular orbital, electrons interact through a variety of mechanisms. 
The first is overall Coulomb repulsion, which we will associate later below with the
Hubbard parameter $U$. The second is Coulomb exchange energy, minimized when the 
molecular state has the highest total spin, and the highest total orbital angular momentum
compatible with it (Hund's rules)~\cite{landau_lifshitz}. The third is the
Jahn-Teller (JT) interaction, caused by coupling of the electron levels to symmetry lowering 
molecular distortions~\citep[chap.~13]{landau_lifshitz}. Contrary to Hund's rules, in a
JT distorted molecule the ground state maximizes double occupancy of levels, thus
favoring low total spin instead of high spin in the isolated molecular ion. 
In molecular C$_{60}^{3-}$, the strength of these interactions has been evaluated 
in the past, and the JT strength has been estimated to prevail narrowly over Hund's 
rule exchange~\cite{Manini02}.  This narrow balance favors a low-spin ground state, 
with a relatively small ``spin gap''-- the energy between the low spin ground 
state and the lowest high-spin excited state, expected to be of the order of 0.1~eV~\cite{PRL01,Manini02}. 
In agreement with this expectation, local moments indicate that in antiferromagnetic 
Mott insulating NH$_3$K$_3$C$_{60}$ the  $C_{60}(3-)$ sites are in a {\em low-spin} 
state, $S$ =1/2~\cite{Prassides1999}, their high spin state $S$ = 3/2 lying about 
100~meV higher in energy. A low-spin qualifies the overexpanded fullerides as ``Mott-Jahn-Teller'' 
insulators -- that is, Mott insulators whose sites are in a JT stabilized 
low-spin state (as opposed to a Hund's rule stabilized high-spin state)~\cite{PRB97}. 
Under hydrostatic pressure, NH$_3$K$_3$C$_{60}$ undergoes a 
transition to a  metallic state where superconductivity re-emerges with a rather 
large $T_c$~\cite{Prassides1999}. It is important to note that this superconducting 
phase still belongs to the ``expanded'' family, as signaled by the fact the $T_c$ here 
\emph{increases} further with increasing pressure (reaching 28~K at 14.8~kbar~\cite{Zhou}) 
at variance with non-expanded fullerides, where $T_c$ drops under pressure.  

A lateral, but relevant additional element comes from  {\em tetravalent} 
compounds A$_4$C$_{60}$, which are insulators or near insulators. By comparison
with the trivalent fullerides, the slight reduction of band-energy gain per particle 
caused by adding one more electron per molecule and by slightly changing the crystal 
structure is sufficient to turn the trivalent metals into tetravalent insulators 
even in  non-expanded materials. Careful density-functional electronic structure 
calculations indicated that it is not possible to describe the tetravalent compounds
such as K$_4$C$_{60}$ as statically distorted Jahn-Teller band insulators~\cite{PRB2000}.  A 
static JT distortion and the associated orthorhombic state is actually present only 
in Cs$_4$C$_{60}$~\cite{Dahlke&Rosseinsky}, and in Rb$_4$C$_{60}$ above a 
critical pressure~\cite{stephens}, while it never shows up in K$_4$C$_{60}$~\cite{stephens} 
(with the exception of monolayers, see \textcite{Crommie}). 
The persistence of insulating or near insulating behavior and the recovery of 
molecular symmetry observed in the high temperature phase of tetravalent fullerides
suggests that these compounds too are Mott-Jahn-Teller insulators~\cite{PRB2000,Knupfer,Kamaras}, 
like the overexpanded trivalent materials. The dynamic JT effect in  each 
C$_{60}^{4-}$ ion associated with Mott localization of carriers is crucial 
in explaining the low spin ground state and the spin gap of A$_4$C$_{60}$, 
exactly as for the expanded trivalent compounds.

From the above discussion one might be tempted to conclude that strong 
correlations play a role only in tetravalent and expanded trivalent compounds, 
while face-centered cubic (f.c.c.) K$_3$C$_{60}$ and Rb$_3$C$_{60}$, where superconductivity 
was originally discovered, could still be viewed as weakly correlated systems, and
as BCS type superconductors. We do not believe in this conclusion.
A final, independent and strongly unconventional signal is provided by NMR. 
In fact, NMR data show direct evidence of a spin gap of order 0.1 eV, appearing as
an anomalous activated increase of inverse relaxation time. 
Most likely this gap between a low spin ground state and a high spin
excited state reflects the multiplet behavior of the localized C$_{60}^{n-}$ molecular ion. 
It shows up ubiquitously in all alkali doped fullerides, 
including superconducting f.c.c. compounds~\cite{Zimmer,Brouet,Brouet-1}. 
The existence of the spin gap signifies that the magnetic response 
of fullerides is very far from Fermi-liquid behavior, which has no such feature.
Magnetically, the fullerides behave as if localized molecular multiplet excitations 
coexisted with delocalized propagating quasiparticles. As discussed, the recovery 
of molecular physics is characteristic of Mott insulators, 
suggesting that the fingerprint of Mott physics is strongly present already 
in the non-expanded superconducting f.c.c. compounds. This suggests
that the f.c.c. compounds are somehow the analog of the overdoped cuprates, 
whereas the expanded trivalent materials are analogous to the underdoped cuprates. 
The conclusion is that both are crucially, even if differently, 
influenced by electron correlations. 

We believe that the above elements are strong enough to call for a 
new physical picture for the whole family of A$_3$C$_{60}$ superconductors. 
Proximity of the Mott insulator strongly suggests that the anomalies of expanded 
fullerene superconductors most likely originate from strong repulsive electron correlation 
in the narrow $t_{1u}$ bands. The prevalence in the Mott localized state of molecular 
physics, with its orbital degeneracy, JT effect and intra-molecular exchange 
must be taken into account along with itinerant electron band physics. Upon 
expanding the cell volume, the intermolecular hopping of electron weakens, whereas 
all the on-site correlation terms -- Coulomb and exchange electron-electron interactions 
as well as molecular JT effect -- are likely to become increasingly relevant. We are 
led to a picture where the Mott insulator physics of weakly coupled molecular 
ions progressively prevails over band physics for increasing lattice expansion. 
In particular, superconductors that operate in this regime are bound to 
deviate from the standard Migdal-Eliashberg scenario, the more so as 
the lattice spacing increases. To investigate that regime, we need to start 
with a broader theoretical scheme for trivalent fullerides, capable of 
describing their behavior under lattice expansion and near the Mott transition. 
While that has been the scope of our work for several years, previous work
was for practical technical reasons limited to tetravalent systems 
~\cite{PRB2000,PRL01,Science}. The study of A$_3$C$_{60}$ systems, computationally 
much heavier due to the simultaneous relevance of magnetic and orbital 
ordering, has only now been completed, and we offer here an outline
of the main results.

\section{Model and Interactions}
Our theoretical model of trivalent fullerides assumes a lattice of 
molecular sites, each representing a C$_{60}$ molecule. The C$_{60}$ $t_{1u}$ 
threefold degenerate LUMO can for all purposes be treated
as an atomic $p$ level. An average of three electrons per molecule are donated by 
alkali atoms and partially fill these orbitals, which can host up to six electrons.
The electrons hop from site to site giving rise to half-filled bands of width 
$W \sim 0.6 \mathrm{eV}$. On each site the electrons experience a Hubbard 
repulsion $U \sim 1\mathrm{eV}$ (corresponding to the Slater integral $F_0 \propto U$),
\be
\mathcal{H}_U = \frac{U}{2}\,\left(n-3\right)^2,\label{H-U}
\ee 
together with a weaker, yet crucial, inter-orbital Hund's rule exchange coupling term
$J_H$ proportional to the Slater integral $F_2$. Under the sole assumption 
of full rotational orbital symmetry, this exchange term takes the form~\cite{PRL01,Science}
\be
\mathcal{H}_{J} = J\,
\left(2\mathbf{S}\cdot\mathbf{S} + \frac{1}{2}\mathbf{L}\cdot\mathbf{L}
\right) + \frac{5}{6}\,J\,\left(n-3\right)^2,
\label{H_el-ph}
\ee
where $J=-J_H<0$, while $n$, $\mathbf{S}$ and $\mathbf{L}$ are the density, 
spin and orbital angular momentum operators, defined as for $p$-orbitals, 
on the given site. This exchange term favors high $\mathbf{S}$ and $\mathbf{L}$ 
molecular multiplets, and has been overlooked or neglected in most treatments 
so far. The next interaction is the JT intra-molecular coupling of electrons 
in the $t_{1u}$ orbital to $H_g$ intra-molecular vibrations. This interaction has
on the contrary been very widely discussed (see e.g.~\textcite{Schluter,Varma,Manini94,
Gunnarsson1995} and references therein), and we will not dwell too much on its details 
here. It acts to split the  $t_{1u}$ orbital degeneracy and thus favors low 
spin states, effectively playing the opposite role 
to intra-molecular exchange. A proper treatment of the JT coupling involves 
the dynamics of carbon nuclei in C$_{60}^{n-}$ ions, including retardation, and
has been developed by \textcite{Gunnarsson1995} 
However for expanded fullerides close enough to the Mott transition, retardation
is not essential and can be omitted. In fact, in this regime the inter-site motion of quasiparticles is severely slowed down~\cite{DMFT} and the coherent bandwidth 
will eventually drop from $W$ to $ZW$ ($Z \ll 1$). When $ZW$ falls enough to
approach the relatively high $H_g$ vibration frequencies of fullerene 
$\hbar\omega \sim 90$~meV~\cite{PRB2000,PRL01,Science},
quasiparticles begin to move on a comparable time scale with the vibrating carbon atoms, and 
non-adiabatic effects become important~\cite{Pietronero}. Even closer to the Mott transition, the
phonon dynamics becomes eventually {\it faster} than inter-molecular quasiparticle hopping. 
In this anti-adiabatic limit, molecular vibrations can be integrated 
away, and the resulting unretarded effective JT interaction recovers identically the same 
form as Hund's rule exchange (\ref{H_el-ph}) except for the sign of $J$, namely
$J_{JT} >0$~\cite{PRL01}. In this limit the Hund and JT intra-molecular inter-orbital 
interaction terms can be directly combined in the form (\ref{H_el-ph}) 
with $J= -J_H + J_{JT}$. In $C_{60}$, $J_H \simeq 0.03-0.1~\mathrm{eV}$~\cite{Martin,Manini02}, 
whereas $J_{JT} \simeq  0.06-0.12~\mathrm{eV}$~\cite{Schluter,Varma,Gunnarsson1995,Manini94,Manini02}.
The total result -- and the only one compatible with s-wave superconductivity, 
with a spin 1/2 Mott insulator, and with a moderate spin gap near 0.1 eV --
is a relatively weak unretarded attraction which has the form of an  
{\em inverted} Hund's rule coupling~\cite{Science,Granath}. This is the
approximation we shall adopt, keeping in mind that it is strictly valid only when 
$ZW < \hbar\omega$, that is, close enough to the Mott transition. 

The lattice expansion characterizing the expanded fullerides is believed to
have little effect on either $ J_H$ or  $J_{JT}$. On the other hand, expansion will surely 
decrease $W$ and increase $U$, so it can be modeled as a gradual increase of $U/W$, 
reflecting both the band narrowing due to smaller overlap between molecular 
wave functions, and a decreased screening strength.  
As discussed above, this Hamiltonian model, even if not really simple, has many body 
interactions that are strictly on site, an ideal situation for Dynamical Mean-Field 
Theory (DMFT)~\cite{DMFT}, one of the most popular and powerful tools in the field of 
strongly correlated electron systems that we briefly describe in the next subsection.

\subsection{Dynamical Mean-Field Theory}
\label{Sec:dmft}
DMFT is a quantum version of classical mean-field theory, which provides 
an exact description of the local dynamics, at the price of freezing away all spatial fluctuations.
The mean-field scheme is formulated by a mapping of the lattice model onto an Anderson 
impurity model (AIM) embedded in a free-electron Fermi ``bath'' subject to a self-consistency condition~\cite{DMFT}.
In our model, the effective AIM is three-fold
orbitally degenerate, with p-like levels representing the $t_{1u}$ orbitals,
each hybridized with a bath. The Hamiltonian is 
\begin{equation}
\label{aim}
{\cal H} = {\cal H}_U +{\cal H}_J +\sum_{ka\sigma} \varepsilon_{ka} c^{\dagger}_{ka\sigma}c^\dagga_{ka\sigma}+\sum_{ka\sigma} 
V_{ka} (c^{\dagger}_{ka\sigma}p^\dagga_{a\sigma} + H.c.),
\end{equation}
where ${\cal H}_U$ and ${\cal H}_J$ are (\ref{H-U}) and (\ref{H_el-ph}) for fermions on the impurity orbitals, 
$\varepsilon_{ka}$ are the bath energy levels labeled by the index $k$ and by an orbital index $a=x,y,z$, 
$V_{ka}$ are the hybridization parameters between the bath fermions, created by $c^{\dagger}_{ka\sigma}$, 
and the impurity fermions, created by $p^{\dagger}_{a\sigma}$.
The mean-field scheme implies a self-consistency condition that depends on the 
impurity Green's function and on the bare density of states of the original lattice. 
Throughout our calculations, we use an infinite-coordination Bethe lattice, 
whose density of states is semicircular. This is a reasonable description of a 
realistic three-dimensional density of states devoid of accidental features 
such as van-Hove singularities. It is moreover particularly convenient since it leads 
to a very simple and transparent form of the self-consistency condition.
The Bethe lattice is the $z\to\infty$ limit of a Cayley tree of coordination 
$z$, scaling the nearest-neighbor hopping in each one of the $z$ directions as 
$t/\sqrt{z}$. The resulting semicircular density of states has bandwidth $=4t$.
  
The self-consistency condition requires the so-called Weiss field 
(i.e., the non-interacting Green's function of the AIM)
\begin{equation}
\label{hybridization}
{\cal G}_0^{-1}(i\omega_n)_a=(i\omega_n+\mu) - \sum_k\frac{V_{ka}^2}{i\omega_n-\varepsilon_{ka}},
\end{equation}
to be related to the local interacting Green's function, $G_a(i\omega_n)$, not only 
through the Dyson equation for the AIM
\[
{\cal G}_0^{-1}(i\omega_n)_a = G_a^{-1}(i\omega_n) + \Sigma_a(i\omega_n),
\]
which requires the full solution of the impurity model, but also by the additional 
selfconsistency equation
\begin{equation}
\label{selfconsistency}
\mathcal{G}_{0a}^{-1}(i\omega_n)=(i\omega_n+\mu) -t^2G(i\omega_n)_a.
\end{equation}

DMFT for a given model thus amounts to solve iteratively the AIM until the impurity Green's function satisfies 
Eq.~(\ref{selfconsistency}). In this work, we solved the threefold degenerate AIM by exact diagonalization.
That requires truncating the sum over $k$ in Eqs.~(\ref{aim}) and (\ref{selfconsistency}) to a finite 
and relatively small number of baths $N_b$, so that the Hamiltonian can be 
diagonalized in the finite resulting Hilbert space. We generally used $N_b=4$ for each orbital.
Most of the results we shall present are at zero temperature, where we can use the 
Lanczos algorithm to calculate the Green's function without fully diagonalizing the Hamiltonian. 
The finite temperature results for the specific heat and its jump at $T_c$ reported in 
Sec.~\ref{Sec:Comparison} are the exception. They are obtained by means of the 
finite-temperature extension of Lanczos~\cite{caponelanczos},
where the thermal Green's function is expressed as a sum over the low-lying eigenvectors 
$\vert n\rangle$ and eigenvalues $E_n$ of the impurity model
\begin{equation}\label{G_sum_Gm}
G_{a\sigma}(i\omega_n)=\frac{1}{Z}\sum_m e^{-\beta E_m} G_{a\sigma}^{(m)}(i\omega_n)
\end{equation}
with
\begin{equation}\label{G_m}
G_{a\sigma}^{(m)}(i\omega_n)\equiv \sum_n
\frac{\left\vert \langle n\vert p_{a\sigma}\vert m\rangle\right\vert ^2}{E_m-E_n-i\omega_n} +
\sum_n \frac{\left\vert \langle  n\vert p^{\dagger}_{a\sigma}\vert m\rangle\right\vert ^2}
{E_n-E_m-i\omega_n}\,.
\end{equation}
The Boltzmann factor in Eq. (\ref{G_sum_Gm}) guarantees that the lower the temperature, 
the smaller the number of excited states 
that actually contribute to the Green's function. Therefore, for sufficiently low temperature, 
we can still use the Lanczos algorithm to find the lowest energy eigenstates.
In practice, only a limited number of states, $20\div 25$, can be calculated in 
a reasonable time. Unfortunately, this number is insufficient to 
make the truncation error in Eq.~(\ref{G_sum_Gm}) negligible for the present model 
in the relevant correlated regime. We estimate the systematic error on the 
Green's function to be of order of a few per cent, a level of accuracy that does 
not allow to fully determine thermodynamic properties such as the critical temperature. 
Luckily though, this error affects much more the absolute specific heat value than 
its relative changes, including the superconducting jump in units of $T_c$ 
which we shall discuss further below (see  Sec.~\ref{Sec:Comparison}). We underline that 
this limitation does not affect by any means the $T=0$ calculations.

The above DMFT equations refer to a paramagnetic phase where no symmetry breaking occurs. 
However, in this work we shall be crucially interested in $s$-wave superconducting 
and in antiferromagnetic solutions, where symmetry is broken.
Superconductivity is conveniently studied in the Nambu-Gor'kov representation 
by introducing spinors
\[
\psi_{\bk a} = 
\left(\begin{array}{c}
p^\dagga_{\bk a\uparrow}\\
p^\dagger_{-\bk a\downarrow}
\end{array}\right),
\]

and defining accordingly the Green's function $G_{\bk a}(\tau) = -\langle T_\tau
\left(\psi^\dagga_{\bk a}(\tau)\,\psi^\dagger_{\bk a}(0)\right)\rangle$ in imaginary time as a $2\times 2$ matrix 
that satisfies the Dyson equation in Matsubara frequencies

\begin{equation}
\label{dyson}
G_{\bk a}(i\omega) = G^0_{\bk a}(i\omega)+ G^0_{\bk a}(i\omega)\,\Sigma(i\omega)\,G_{\bk a}(i\omega),
\end{equation}
with $G^0_{\bk a}(i\omega)$ the non-interacting value. The single-particle self-energy $\Sigma(i\omega)$ 
is also a $2\times 2$ matrix whose off-diagonal element $\Delta(i\omega)$, when finite,  
signals a superconducting phase. The DMFT self-consistency can be written now as
\begin{equation}  
\label{selfsuper}  
\mathcal{G}_0^{-1}(i \omega_n)= i\omega_n \,\tau_0 + \mu \,\tau_3 
 -t^2 \, \tau_3 \,G(i\omega_n)\, \tau_3,
\end{equation}  
where $\tau_{0}$ and $\tau_{3}$ are Pauli matrices. 

Analogously, antiferromagnetism in a bipartite lattice is conveniently described using the spinor 
\[
\psi_{\bk a \sigma} = 
\left(\begin{array}{c}
p^\dagga_{\bk a\sigma}\\
p^\dagga_{\bk + \bQ a\sigma}\end{array}\right),
\]
with $\bk$ in the magnetic Brillouin zone and $\bQ$ the modulation vector. 
This leads once more to a $2\times 2$ Green's function and self-energy matrices related by the Dyson equation 
(\ref{dyson}). Here too, a finite off-diagonal element signals an antiferromagnetically ordered phase.
The self-consistency equation for antiferromagnetism exploits the bipartite property of the lattice. 
Indicating one sublattice with A and the other with B, the general self-consistency condition is  
\begin{equation}
\label{selfconsistencyafm}
{\cal G}_0^{-1}(i\omega_n)_{\sigma A}=(i\omega_n+\mu)-t^2\,G(i\omega_n)_{\sigma B},
\end{equation}
which becomes Eq.~(\ref{selfconsistency}) if $G_{\sigma A} \equiv G_{\sigma B}$, i.e. if 
the system is nonmagnetic. When the system is antiferromagnetic, then $G_{\sigma A}\equiv G_{-\sigma B}$. 
Thus we can eliminate the sublattice B from Eq. (\ref{selfconsistencyafm}), and obtain the following 
result for the self-consistency equation: 
\begin{equation}
\label{selfconsistencyafm-bis}
{\cal G}_0^{-1}(i\omega_n)_{\sigma A}=(i\omega_n+\mu)-t^2G(i\omega_n)_{-\sigma A}.
\end{equation}
This equation is valid for a Bethe lattice with nearest neighbor hopping, a case with 
perfect nesting that is rather exceptional in realistic antiferromagnets. 
A way to simulate imperfect nesting typical of more realistic situations, while still taking 
advantage of the Bethe-lattice simplifications, is to add a next-nearest-neighbor 
hopping $t^\prime/z$ in the Cayley tree (merely a device to eliminate nesting, 
not meant to suggest next nearest neighbor hopping, small in fullerides). In the limit 
$z\to\infty$ of the Bethe lattice, the self-consistency equation becomes
\begin{eqnarray}
{\cal G}_0^{-1}(i\omega_n)_{\sigma A}&=& (i\omega_n+\mu)-t^2\,G(i\omega_n)_{\sigma B}-{t^{\prime}}^2\,G(i\omega_n)_{\sigma A}
\nonumber\\
&=& (i\omega_n+\mu)-t^2\,G(i\omega_n)_{-\sigma A}-{t^{\prime}}^2\,G(i\omega_n)_{\sigma A}.
\label{selfconsistency2}
\end{eqnarray}

For both broken-symmetry phases, if the diagonal elements of the self-energy matrix 
at low Matsubara frequencies  follow the conventional Fermi-liquid behavior -- which 
we always find to be the case,  
\begin{equation}
\Sigma_{diagonal}(i\omega_n) \simeq \left(1-\frac{1}{Z}\right)\,i\omega_n,
\label{zeta}
\end{equation}
the actual value of the spectral gap in the single particle spectrum is given by 
$Z\,\Delta(0)$, the zero frequency anomalous (superconducting or antiferromagnetic) 
self-energy multiplied by the so-called quasiparticle weight $Z$.

We end by noting that, within DMFT, one can search for solutions with different 
symmetries by simply allowing/preventing symmetry breaking order parameters, even 
in regions where the chosen phase is not the most stable. When two or more 
solutions coexist, the stable one is determined by an explicit energy 
calculation. As we shall discuss, for a wide range of $U/W$ values 
we do find coexisting superconducting and antiferromagnetic solutions,
the former prevailing at smaller $U$, the latter at larger $U$. 
The physical phase diagram thus exhibits a first-order transition 
between these two symmetry broken phases -- a nonmagnetic $s$-wave superconductor,
and an insulating spin 1/2 antiferromagnet -- taking place when 
the respective energy curves intersect.

\subsection{$T =$ 0 Phase Diagram}
Modeling lattice expansion of fullerides as a gradual increase of $U/W$, 
we can proceed to analyze the theoretical zero temperature ``phase diagram'' of our model
obtained by DMFT as a function of $U/W$.
Starting with the uncorrelated system with $U = 0$ the model initially exhibits 
straight BCS superconductivity driven by JT phonons 
(i.e., the attractive $J$ we introduced above) with an $s$-wave ($S = L = 0$) 
order parameter\footnote{Since gauge symmetry is broken, we are allowed to assume $P_{SC}$ real}  
\be
P_{SC} = \frac{1}{N}\sum_i\sum_{a=x,y,z}\,
\langle p^\dagger_{i\,a\uparrow}\, p^\dagger_{i\,a\downarrow}\rangle,
\label{Cooper-channel}
\ee
where $N$ is the number of molecules and $p^\dagger_{i\,a\sigma}$ creates an electron on molecule $i$, with spin $\sigma$ and in orbital   
$a=x,y,z$.  The Fermi-liquid scattering amplitude in 
the Cooper channel, measuring the strength of
the effective attraction, is $A=-10 J/3$. We note that owing to fairly strong JT interactions
~\cite{Manini94, Gunnarsson1995}, if Hund's exchange $J_H$ were neglected, the
dimensionless JT coupling constant of fullerene controlling superconductivity would be 
numerically very large, $\lambda = \rho_0\vert\,A\, \vert\simeq 0.6 - 1.0$, where 
$\rho_0$ ($\simeq 2.4~\mbox{eV}^{-1}$) is the bare density of states per spin and orbital.
Turning on a weak Coulomb repulsion $U$ on top of that will reduce the pairing attraction 
in this regime. Perturbatively one obtains for small $U$ that $A= -10\, J/3 + U$. 
Since $J$ is insensitive to expansion while $U/W$ increases, this implies 
that in this picture, where Hund's rule exchange is neglected, $T_c$ should 
always {\it decrease} upon expansion, a prediction which is at odds with experiments. 

In fact, as anticipated, Hund's rule exchange is not negligible, and its 
effect is to introduce a substantial cancellation in $J$ leading to a largely
reduced effective coupling $\lambda_{eff} \simeq \frac{10}{3}\left(J_{JT}-J_H \right)\,\rho_0 $.
Should we interpret the ubiquitous spin gap of 0.1~eV~\cite{Zimmer,Brouet,Brouet-1} as 
the C$_{60}^{n-}$ molecular excitation energy between its low spin ground state 
with $S= 1/2$ and $S=0$ for $n=3$ and $n=4$, respectively, and its high spin excited 
state, with $S= 3/2$ and $S=1$ for $n=3$ and $n=4$, respectively, we would conclude 
that, to be consistent with our model where both gaps are equal to $5J$~\cite{Capone04},   
the total effective inverted exchange $J$, comprehensive of both JT and Hund's rule
exchange, should be approximately $J=0.02$~eV. In reality, the 
qualitative scenario we will describe is relatively independent of the precise 
value of $J$ provided it is inverted, i.e., negative and small. 

Once exchange is included, the resulting $\lambda_{eff} \simeq 0.16$ is now much 
smaller -- in fact way too small to explain within conventional BCS or Migdal-Eliashberg theory  
any of the observed values of $T_c$ in fullerides (let alone the non-monotonic behavior 
of $T_c$ versus the density of states for expanded fullerides~\cite{Dahlke,Ganin}). 
Things get worse when we increase the on-site Coulomb repulsion $U$ closer to
realistic values, $U\sim W$ and beyond. In the conventional weakly correlated picture 
$U$ would provide in the electron pairing problem a repulsive ``Coulomb pseudopotential'' whose 
bare value is $\mu_* = U\,\rho_0\simeq 3$~\cite{BCSreview}. Simply comparing 
these bare values of $\lambda$ and $\mu_*$ we should conclude that $s$-wave BCS 
superconductivity  in fullerenes is simply impossible (with the obvious proviso that for small $U$ 
an unretarded treatment of JT phonon interactions is not really justified).

\begin{figure}
	\includegraphics[width=8.5cm]{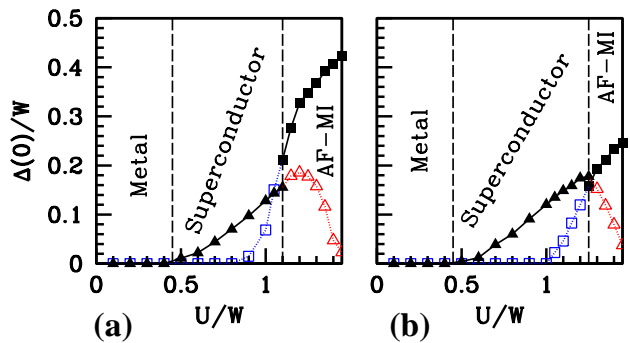}
	\caption{\label{phase-diagram} Superconducting solution (triangles) and antiferromagnetic solution (squares) 
	zero-frequency anomalous self-energies $\Delta(0)$ as function of $U/W$. Solid symbols are used when the 
        corresponding symmetry broken phase is stable, while open symbols when it is metastable, i.e. has lower energy 
        then the other phase.  The first-order transition between the two phases is indicated by a 
        vertical line separating the superconductor from the antiferromagnetic Mott insulator. 
        Panel (b) corresponds to the case in the presence of a frustrating next-nearest neighbor 
        hopping $t'=0.3 t$, absent in (a).}
	\end{figure}

The full DMFT solution of the model for $J=0.05~W$ and increasing $U/W$  
yields the phase diagram in Fig.~\ref{phase-diagram}. While confirming the above
expectations for moderate $U$, it has a surprise in reserve at 
larger $U$ values, where the Mott transition is approached.

\begin{figure}
\includegraphics[width=8.5cm]{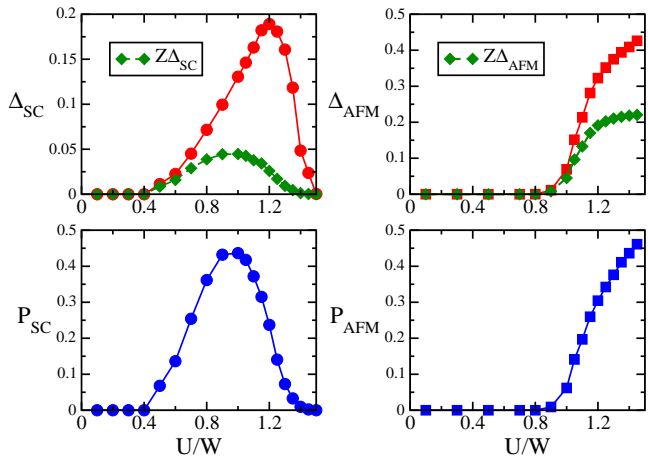}
\caption{\label{opdelta} Superconducting solution (SC, left) and antiferromagnetic solution (AFM, right) 
anomalous self-energies (top) and order parameters (bottom) as a function of $U/W$. 
The top panels also show (green diamonds) the spectral gaps obtained multiplying $\Delta$ by the quasiparticle 
weight $Z$ of each solution. The gaps are in units of the bandwidth $W$ (the order parameters are 
by definition dimensionless). Dashed vertical lines mark the first-order 
phase transition between the two solutions.}
\end{figure}

Fig.~\ref{opdelta} shows the zero frequency anomalous single-particle self-energies calculated within DMFT 
for the superconducting and for the antiferromagnetic solution. At $U=0$, the model is 
an $s$-wave BCS superconductor, with an exponentially small superconducting $\Delta(0)$. 
It is too small to be visible in the figure, since the effective exchange-reduced $\lambda \sim 0.2$ 
is as was said very weak. Beginning from zero, the increase of $U$ first rapidly 
destroys the weak BCS superconductivity. The superconducting $\Delta(0)$ vanishes at 
roughly the mean-field value $U=10/3 J$, and above this value of $U$ the ground state 
becomes a normal metal as expected. Upon further increasing $U/W$ the model remains 
a normal metal -- no superconductivity, no antiferromagnetism. However, the importance 
of electron correlations increases with $U$, as signaled for example in the DMFT 
spectral function (not shown) by the gradual
formation of incoherent Hubbard bands on both sides of the Fermi level. The metallic 
character persists until a hypothetically continuous Mott transition eventually 
reached near $U/W \sim 1.5$, where $Z=0$ and the metallic character is extinguished. 

Before this happens however, $s$-wave superconductivity re-enters from the normal metal
state. The anomalous self energy $\Delta(0)$, proportional to the superconducting 
order parameter has, as a function of $U$, a bell-shaped behavior --
a ``superconducting dome'' as it is called in cuprates -- hitting a large maximum before dropping again. The 
re-entrant superconductive behavior is a clear realization of phonon-induced strongly 
correlated superconductivity (SCS)~\cite{Science}. The sharply rising order parameter
edge with increasing $U/W$ can in our view explain the strong rise of $T_c$ upon 
lattice expansion in non-expanded compounds, previously (and we believe incorrectly)
attributed to a BCS-like increase of density of states upon band narrowing. 
Past the dome maximum, and upon increasing expansion, the SCS superconducting order
parameter declines, and would eventually drop to zero at the continuous metal 
insulator transition near  $U/W \sim 1.5$. This continuous decline of superconductivity
is preempted by a first order transition to a lower energy antiferromagnetic 
Mott insulating phase, with order parameter
\begin{equation}
P_{AFM}=\frac{1}{N}\sum_{i}(-1)^i(n_{i\uparrow}-n_{i\downarrow}),
\end{equation}
where $n_{i\sigma}=\sum_{a=x,y,z} p^\dagger_{i\,a\sigma}p^\dagga_{i\,a\sigma}$ is the full occupation number 
with spin $\sigma$ at molecule $i$. 
The exact location of the superconductor-insulator transition depends on details. 
For strong nesting ($t^\prime$=0) it takes place even before the superconducting dome maximum. 
In Fig.~\ref{opdelta} we show $\Delta(0)$ of Fig.~\ref{phase-diagram} 
in comparison with the spectral gaps $Z\Delta(0)$, as well as the order 
parameters $P_{SC}$ and $P_{AFM}$.  Notice that $Z$ for the superconductor is smaller and vanishes at the continuous  
metal insulator transition, while $Z$ for the antiferromagnet is of order 1. In both 
cases the dimensionless order parameter essentially follows the behavior of the spectral gap.

When we add a next-nearest neighbor hopping $t^\prime$ to mimic imperfect nesting 
(as we expect to find generically for realistic band structures,  
in particular in the face-centered cubic A$_3$C$_{60}$ materials), 
we find that the superconducting phase is only weakly affected but antiferromagnetism is 
strongly frustrated. As a result the superconducting region expands at expense of the 
magnetic insulator, and the superconducting dome may emerge in full 
[panel (b) of Fig.~\ref{phase-diagram}]. We propose that the gradual drop of superconducting
order parameter past the dome maximum now naturally explains the decline of $T_c$
of expanded fullerides~\cite{Dahlke&Rosseinsky, Durand2003,Ganin}.

Finally, past the first-order Mott transition, we find that the antiferromagnetic insulator 
is formed mainly by spin-1/2 local configurations, which is in agreement with experiments 
in NH$_3$K$_3$C$_{60}$~\cite{Iwasa}. We also predict that ambient pressure 
A15 Cs$_3$C$_{60}$, yet to be characterized, should similarly be a spin-1/2
antiferromagnetic insulator. Besides spin rotational symmetry, this kind of
state also breaks orbital rotational symmetry, signaling that spin ordering 
must be generally accompanied by orbital ordering. In ammoniated fullerides that
again is consistent with experiment~\cite{Iwasa}. We conclude that, in spite of strong 
simplifying assumptions, our model seems able to reproduce very important features of the 
phase diagram of expanded fullerides. In the following we shall discuss in more
detail the strongly correlated superconducting phase near the Mott transition, and 
also propose experiments that might distinguish it from a standard BCS state.

\section{Understanding Strongly Correlated Superconductivity from DMFT and the Impurity Model}
The re-emergence of phonon-driven superconductivity close to the Mott transition -- 
Strongly Correlated Superconductivity (SCS) -- was discussed in \textcite{Science} 
in terms of Fermi-liquid theory. A key point of that phenomenon is the renormalization of 
the effective bandwidth and thus of the effective mass, both controlled (in a Bethe 
lattice) by the  quasiparticle weight $Z$, Eq.~(\ref{zeta}). $Z(U)$ decreases as a function of
$U/W$ and vanishes at the continuous metal insulator transition point $U=U_c$, where the effective 
mass $m^*/m = 1/Z(U)$ diverges, and the effective quasiparticle bandwidth $W_* = ZW$ vanishes.
An estimate of the interaction between charged quasiparticles requires the evolution of fluctuations that 
take place in charge space. Because charge fluctuations are gradually frozen away near the Mott transition, 
the effective repulsion between quasiparticles is also renormalized down to some smaller value $U_*<U$. 
In particular, the Fermi-liquid description provided, e.g., by the Gutzwiller variational 
approach~\cite{gutzwiller} and supported by the DMFT behavior of the average charge 
fluctuations $\langle (n-3)^2\rangle$, suggests that $U_*\simeq U\,\Gamma_U\,Z(U)^2 \sim U\,Z(U)$, where 
$\Gamma_U$ includes all the so-called vertex corrections~\cite{abrikosov}. This implies that  
the vertex function $\Gamma_U$ diverges close to the Mott transition, but does not compensate the vanishing of $Z$~\cite{Science}. 
The pairwise Jahn-Teller and exchange based attraction $J$ between quasiparticles, 
even if small, here is restricted to spin and orbital space, and has nothing to do with charge
fluctuations. In other words, Hund's rule exchange and the JT coupling only influence 
the internal splitting of each molecular multiplet without affecting its center of gravity.  
As a result, this attraction should remain unrenormalized $J_* = J\,\Gamma_J\,Z(U)^2 \sim J$ close to the 
metal insulator transition, thus implying a strongly divergent vertex correction $\Gamma_J$ that 
cancel the vanishing $Z$. Therefore the electron pair scattering amplitude $A_*$ in the
Cooper channel should renormalize as $A = U - \frac{10}{3}J \rightarrow A_* = Z(U)\,U - \frac{10}{3}J$.
When $U/W$ is small, $Z \lesssim 1$, the main effect of $U$ is to suppress 
superconductivity, as was noted earlier. However, if $U$ is close to the critical 
metal-insulator value $U_c$,  then $Z \sim (U_c-U)/U_c\ll 1$ and the scattering amplitude 
turns negative in spite of a large $U$. This is qualitatively the reason for the SCS re-entrance 
of superconductivity (though in this region of course the actual pair scattering amplitude 
might deviate from this simple formula~\cite{Science}). 

\begin{figure}[t]
\includegraphics[width=8cm]{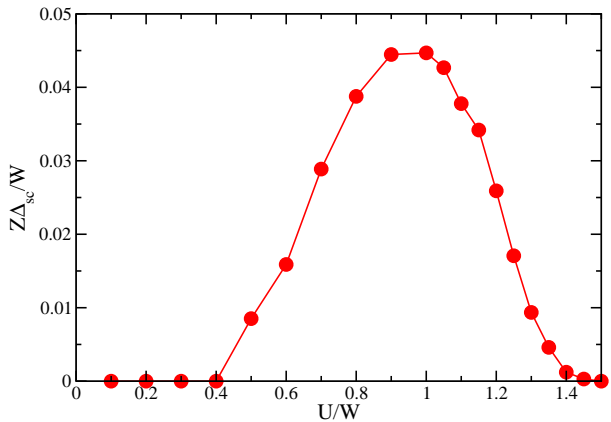}
\caption{\label{gap} Quasiparticle superconducting energy gap in units of the bandwidth $W\simeq 0.6$~eV and on a larger scale than 
Fig.~\ref{opdelta}, computed through 
the anomalous self-energy, panel (a) in Fig.~\ref{phase-diagram}, multiplied by 
the quasiparticle residue $Z(U)$. Notice that, above $U/W\simeq 1.1$, 
the superconducting solution is metastable since the antiferromagnetic one has 
lower energy, see Fig.~\ref{phase-diagram}. We note that the maximum gap $Z\,\Delta_c\simeq 0.045~W\simeq 27$~meV.}
\end{figure}

In addition, the Fermi-liquid 
argument suggests an explanation for the large value of superconducting order parameter,
implying a large $T_c$, in the SCS regime, see Fig.~\ref{phase-diagram}, compared to the $U = 0$ BCS 
values. In fact, when $A_* \simeq Z\,W$, and $Z$ is dropping sharply, the quasiparticle 
attraction $A_*$ will at some point $U=U*$ equal the coherent quasiparticle bandwidth $ZW$. That very 
uncommon situation, of metallic quasiparticles with an pair attraction equal to their
energy bandwidth, is known to yield maximum superconductivity for a given attraction.
As shown by studies of purely attractive models~\cite{Robaszkiewicz},  the 
maximum superconducting temperature $k_B T_c$ attainable in that case is about 7\% of 
the pair attraction energy itself. In our model of trivalent fullerides, this estimate yields 
$k_BT_c \sim 0.07 A_* \sim 0.2 |J|$, which has the correct magnitude of roughly 
40 K for $J \sim$ 20 meV -- a value in turn fully consistent with the observed spin gap
0.1 eV $\sim 5 J$. While this coincidence of numbers is probably fortuitous, it does indicate 
that orders of magnitude implied by our model with realistic parameters are quite 
consistent with experimental facts. At face value, it also suggests that 
0.07 $|J|$ = 0.07 $(J_{JT} -J_H)$ could be the maximum attainable $k_BT_c$ in fullerides.
We conclude that strong correlations play a crucial 
role in bringing the superconducting gap magnitude to the right range of values 
as compared with the experimental ones, see Fig.~\ref{gap}. Such values and
large critical temperatures would never be attained within conventional BCS theory 
using a value of $\lambda \simeq 0.16-0.2$, including as it should the large cancellation 
of JT by exchange. They could in point of fact be attained if the cancellation due to 
exchange were (incorrectly) neglected; but then a lattice expansion should always 
lead to a decrease of $T_c$, contrary to experiment. 

\begin{figure}[t]
\includegraphics[width=8cm]{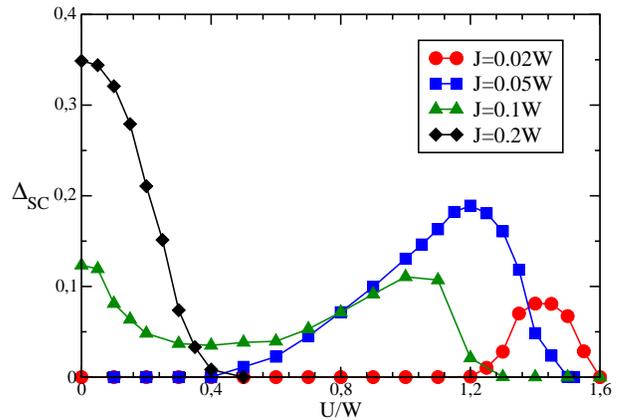}
\caption{\label{VariousJ} Anomalous self-energy $\Delta(0)$ (related to the superconducting 
gap by a factor $Z^{-1}$) for the superconducting solution and different values of the coupling 
parameter $J=0.02$, $0.05$, $0.1$ and $0.2$, which correspond to 
$\lambda=0.09,~0.21,~0.42,~0.85$, respectively. Note that for large $J$, corresponding
to a case where the Jahn Teller coupling is not canceled by Hund's rule exchange, superconductivity
is strongest at $U$ = 0. For increasing cancellation (decreasing $J$), two separate superconducting
pockets emerge, a BCS pocket near $U$ = 0, and a SCS pocket near the Mott insulator. When the 
cancellation is nearly complete, SCS is many orders of magnitudes stronger than BCS. This is the situation we propose in our model of fullerides. Similar physics was described for a simpler model in \textcite{Capone04}.}
\end{figure}

To appreciate further the effect of exchange-JT cancellation, it is instructive 
to consider, as was done for a simplified model by \textcite{Capone04}, 
the behavior with $U/W$ of the superconducting self-energy $\Delta(0)$ (proportional to the 
$T=0$ gap, and roughly speaking to $T_c$) starting with pure JT and without exchange, 
and then proceeding to turn on exchange and gradual cancellation, see Fig.~\ref{VariousJ}.  
For the bare JT,  $\lambda\simeq 1$ (a strong coupling value) the superconducting 
self-energy decreases monotonically with increasing $U$, in agreement with the 
Migdal-Eliashberg prediction of an increasing Coulomb pseudo-potential. Above a 
critical value, the system turns directly, via a second-order or weakly first-order 
phase transition, to a Mott insulating phase. This result is fully consistent with 
previous calculations by Han {sl et al.}~\cite{Han2003}, where the same type of 
Hubbard model was studied within DMFT at finite temperature.
Treating explicitly the electron-phonon coupling (including the full phonon dynamics) 
with $\lambda=0.6$ and neglecting exchange they obtained a superconductor with 
monotonically decreasing gap. 

Through a progressive reduction of $\lambda$ (mimicking JT cancellation by 
exchange) we find that a non-monotonic superconducting behavior makes its appearance
as a function of  $U$.  Initially there is still a single superconducting phase for all 
$U/W$ values; but two different regions near zero and near $U_c$ begin to materialize.
(Note that $U_c$ simultaneously shifts to higher $U$ as $\lambda$ decreases).
When the cancellation is so strong that $\lambda$ is still positive but small, 
the two superconducting regions break apart to form two separate pockets, 
leaving a normal metal phase in between. In the leftmost pocket near $U/W =0$
the anomalous self-energy has a BCS-like exponential dependence on $\lambda$ 
and indeed superconductivity in this corner is BCS. Superconductivity in 
the rightmost pocket near the metal insulator transition behaves quite 
differently. Here the $\lambda$ dependence of superconductivity is much weaker, 
and the superconductive gap much stronger, than in the BCS pocket.  Superconductivity 
in this pocket can in fact be characterized as SCS~\cite{Science}, due to 
narrow quasiparticle pairing as described above. A similar behavior to Fig.~\ref{VariousJ}, 
with two separate BCS and SCS regimes emerging from a single initial one when  
the effective pairing attraction is progressively weakened by exchange, was derived and illustrated 
in a simpler twofold degenerate model in \textcite{Capone04}.

The SCS superconducting pocket near the Mott transition is expected to differ from the BCS pocket
even in its normal state properties.  The normal state underlying a BCS superconductor 
is Fermi-liquid-like. On the other hand, previous analysis suggest that the Fermi-liquid picture 
is likely to break down in our model when the Mott transition is approached. 
The key reason for the breakdown of the Fermi-liquid is precisely that, when 
$Z \rightarrow$ 0, the attraction between quasiparticles must eventually reach and exceed 
in magnitude the quasiparticle bandwidth $ZW$, a situation difficult to sustain.\footnote{Note that the high-energy Hubbard bands are 
unaffected by the small attraction $J$; superconductivity is just a matter of 
quasiparticles~\cite{Science}. Therefore, a quasiparticle attraction exceeding their bandwidth cannot correspond to an instability 
towards a Mott insulator, which must involve also the Hubbard bands, but at most towards a breakdown of a quasiparticle-based Fermi liquid.} 
Possible deviations from the Fermi-liquid paradigm were
in fact overlooked in Ref.~\cite{Science} as they are related to the very-low energy behavior of 
the normal phase close to the Mott transition, not explored in that work. Later, 
the non-Fermi-liquid behavior was discovered in the two-orbital model 
where the physics is very similar~\cite{Capone04}.

\subsection{Anderson Impurity With a Rigid Bath}

\begin{figure}[t]
\includegraphics[width=7cm]{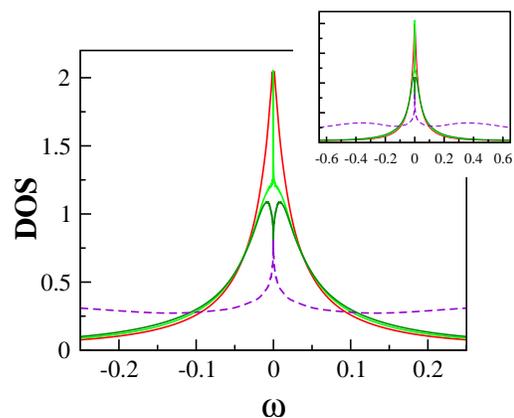}
\caption{\label{DOS} Impurity spectral function with 
a rigid bath characterized by a flat density of states (no DMFT self-consistency) across 
the critical point: the curves from top to bottom 
correspond to the evolution from the Kondo screened regime towards the pseudogap regime. 
The dotted line is the spectral function 
deep inside the non-Fermi-liquid phase $J\gg T_K$. The top inset shows the same curves plotted on a 
larger energy range, where the pseudogap of the dotted line is visible. The Fermi liquid
behavior corresponds to the normal state of the SCS superconducting phase for $J < J_{cp}$ 
(analogous to the overdoped regime of cuprates) whereas the pseudogap behavior 
corresponds to the SCS normal state for $J > J_{cp}$ (analogous to the underdoped cuprates). 
Figure from \textcite{DeleoPRL}.}
\end{figure}

The DMFT calculations described above involved two steps, one solving the
Anderson impurity model (AIM), the other making that selfconsistent with the
bath. Following a reasoning proposed by \textcite{DeLeo2003}, one 
may start off with the first step alone, namely analyzing the bare AIM 
without imposing any self-consistency constraint. The conduction bath
can be assumed to have a flat density of states, and the bath-impurity hybridization
to be structureless, a situation which avoids numerical uncertainties 
and yields accurate low-energy properties. This kind of analysis applied to the  
AIM (\ref{aim}) shows~\cite{DeleoPRL} that two different impurity 
phases are stabilized according to the ratio between the attraction $J$, 
and the Kondo temperature $T_K$~\cite{Kondo}. Below this temperature and when $J=0$, the spin of an impurity
coupled to a Fermi sea is screened out and absorbed in the conduction sea~\cite{Kondo}. 
In the lattice context within DMFT, the Kondo scale measures metallic coherence and  
corresponds to the renormalized quasiparticle bandwidth $ZW$. For finite $J\not =0$ but smaller 
than $T_K$, Kondo screening remains, thus still implying a  Fermi-liquid behavior in DMFT.
In fullerides, the impurity represents the C$_{60}^{3-}$ ion, carrying three
orbitals and three spins. In the Kondo phase each of the three spins is separately 
screened by the bath and thus incorporated in the Fermi sea. 
Conversely, when $J > T_K$ the Kondo screening is lost, and that 
was shown to imply a non Fermi-liquid phase characterized by a pseudogap in the 
single-particle spectrum and by several other singular properties~\cite{DeleoPRL}.  
A very qualitative description of this phase is that, unlike the Kondo phase, 
two spins out of three pair off antiferromagnetically at any given time, leaving 
out a single spin 1/2 available for Kondo screening. However, since orbital degeneracy is unbroken, 
this residual spin $S=1/2$ also carries orbital momentum $L=1$, which corresponds to an overscreened 
non-Fermi liquid situation~\cite{DeleoPRL}. 
In this regime, it was predicted that the impurity contributions
to the specific-heat coefficient and to the pair susceptibility in the $s$-wave 
channel (\ref{Cooper-channel}) diverge as $T^{-1/5}$ at low temperature $T$. 
In addition, the conduction electron scattering rate has a non-analytic 
temperature-behavior $T^{2/5}$. The local response functions to either
a quadrupolar field, which splits the orbital degeneracy, or to a magnetic 
field which polarizes both spins and orbitals also diverge as $T^{-1/5}$.     
The two phases, the Kondo screened phase and the pseudogap phase, are separated 
by a critical point at $J = J_{cp} \simeq T_K$. It is endowed with a finite entropy 
$1/2\,\ln 3$, and with a divergent superconducting susceptibility with an exponent $1/3$. 
As shown in Fig.~\ref{DOS}, the single-particle spectral function displays strong 
deviations from a normal metal in the pseudogap phase and at the critical point.

Near this critical point it has been found that the low-energy dynamics around the impurity is controlled 
by two separate energy scales~\cite{DeleoPRL}, $T_+$ and $T_-$, 
whose behavior is very different as a function of increasing $U/W$. 
A higher energy scale $T_+$ is set by the critical $J_{cp}\sim T_K$ and represents 
the width of a broad incoherent resonance; it evolves smoothly and uneventfully
as the critical point is crossed (See Fig.~\ref{DOS}). A lower energy scale 
$T_-\propto |J-T_K|^3$ measures instead the distance from the critical point, 
and leads simultaneously to a narrow resonance in the Fermi-liquid region, and to 
an equally narrow spectral density dip (the ``pseudogap'') in the pseudogap 
phase~\cite{DeleoPRL}. Since in this phase the impurity still carries a residual 
spin-1/2, there remains a finite value of the spectral function at the chemical 
potential, and the gap is not complete. The pseudogap widens if $J$ is increased,  
and the cusp-like dip in the impurity spectral function smoothly turns into a 
cusp-like peak, the value at the chemical potential staying fixed and constant. 
This behavior is shown by the dotted curve in Fig.~\ref{DOS} corresponding to a 
very large pseudogap (see top inset), possessing a very tiny peak at the 
chemical potential. This indicates the existence of yet another energy scale 
besides $T_+$ and $T_-$ that sets the width of the cusp peak.

\subsection{Anderson Impurity with a Self-Consistent Bath in DMFT}

The rigid bath AIM behavior and its critical point briefly reviewed above 
provide a guide to the DMFT results once the impurity-bath coupling is self-consistently 
determined. First of all, since $T_K$ coincides within DMFT with the renormalized $ZW$ 
which in turn vanishes when the continuous Mott transition is approached, 
the impurity critical point is inevitably met before the metal insulator transition
as $U/W$ is increased, at some $U_{cp} \alt U_c$. This entails several important consequences:

\begin{itemize}
\item The normal state may be a Fermi liquid only far below the continuous metal insulator
transition, such as perhaps may be the case in the non-expanded fullerides. Expanded 
compounds on the other hand are expected to have a non Fermi-liquid normal state, 
and eventually a pseudogap, possibly developing before the first order transition 
to the antiferromagnetic insulator.

\item The SCS superconducting pocket near the Mott transition reflects the leading 
instability of the impurity critical point. In other words SCS superconductivity 
is the way in which the lattice model responds to impurity criticality and avoids it.

\item The lower energy scale vanishes right at the critical point, $T_-=0$. Here 
$T_+\simeq J$ thus remains as the only energy scale controlling the magnitude of the 
superconducting energy gap. Away from $U=U_{cp}$, $T_-\not =0$, and the amplitude of the 
superconducting gap should decrease monotonically with $(T_+-T_-)/T_+$~\cite{Capone04}, since 
$T_-$ cut-offs the local pairing instability. Therefore the gap should be maximum right at the impurity critical point (the top of the dome).

\item Even though the normal phase is non-Fermi liquid, well defined Bogoliubov 
quasiparticles should exist inside the SCS superconducting pocket. 

\end{itemize}
The last statement comes from the fact that, at the impurity critical point, superconductivity 
provides a new screening channel which helps the system get rid of the finite 
residual entropy at the critical point, thus eliminating non-Fermi liquid singularities~\cite{DeleoPRL,SchiroPRB}. 

\begin{figure}[t]
\includegraphics[width=7cm]{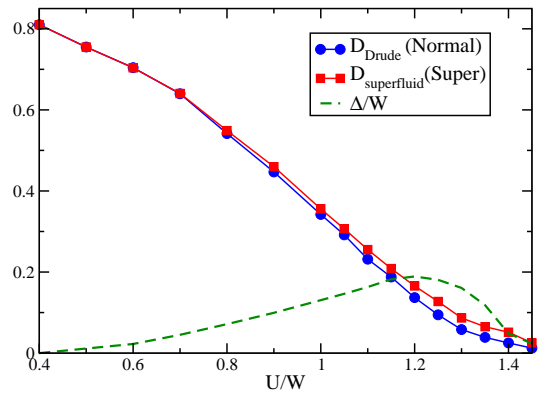}
\caption{\label{Drude} Calculated weights of the zero-frequency delta-function contribution 
to the optical conductivity for the superconducting phase (superfluid stiffness, squares) 
and the normal phase (Drude weight, circles). The zero-frequency anomalous self-energy is also plotted.
Note the higher values in the SCS superconductor, relative to the nonsuperconducting metal phase.}
\end{figure}

\begin{figure}[t]
\includegraphics[width=7cm]{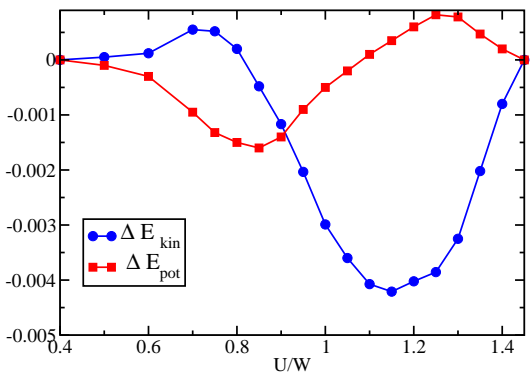}
\caption{\label{kinetic.vs.potential} Energetic balance underlying superconductivity.  $\Delta E_{pot} = E^{S}_{pot} - E^{N}_{pot}$ 
is the difference between the potential energies of the superconducting and the normal solution, 
while $\Delta E_{kin} = E^{S}_{kin} - E^{N}_{kin}$ is the same difference between the kinetic energies of the two solutions.}
\end{figure}

The role of superconductivity as a novel screening channel close to the 
impurity model critical point should reflect, in the lattice model, into  
a gain of band energy (the tight binding ``kinetic energy'') at the onset of superconductivity. 
Owing to a sum rule connecting kinetic energy and zero-frequency optical
conductivity (sometimes referred to as ``Drude weight'')~\cite{Drude}
the onset of SCS superconductivity close to the Mott transition
should lead to a Drude weight increase. This prediction is 
well borne out by the full DMFT solution of our Hamiltonian.
In Fig.~\ref{Drude} we plot the $\omega=0$ (d.c.) optical conductivity of 
our model superconductor (where it coincides with the superfluid stiffness), defined by~\cite{toschi}
\begin{equation}
D_s= -E_{kin} + \chi_{jj}({\bf q} \rightarrow 0, \Omega=0),
\label{ds}
\end{equation}   
where $E_{kin}$ is the kinetic energy and $\chi_{jj}$ the static limit of the paramagnetic
part of the electromagnetic kernel
\begin{eqnarray}
\chi_{jj} & = & \frac{2}{\beta} \sum_{n} \int \, d\epsilon \, N(\epsilon) \, 
V(\epsilon) \nonumber\\
&\times & \left[ G(\epsilon,\omega_n) G^{*}(\epsilon, \omega_n) +
F(\epsilon,\omega_n)F(\epsilon,\omega_n) \right], 
\label{chijj} 
\end{eqnarray}         
where $V(\epsilon)=(4t^2-\epsilon^2)/3$ is the current vertex in the Bethe lattice, while $G(\epsilon,\omega_n)$ and  
$F(\epsilon, \omega_n)$ are  the normal and anomalous lattice Green functions, respectively. 
In the same figure we also plot the zero-frequency d.c. conductivity (the Drude weight) of the  
underlying metastable solution where superconductivity is 
inhibited, this state meant to provide a cartoon of the real normal phase above $T_c$. 
The Drude weight is given by Eq. (\ref{ds}) and 
(\ref{chijj}) with $F(\epsilon,\omega_n)\equiv0$. Upon entering the SCS dome 
from the low $U/W$ side, the superconducting phase initially loses kinetic energy 
over the normal state as in ordinary BCS theory. However, upon increasing $U/W$ at 
and beyond the dome maximum, the loss reverts over to a gain, and indeed 
most of the SCS superconducting pocket is predicted to have a larger 
weight of the zero-frequency optical absorption than the non superconducting state.
The same behavior is displayed (Fig.~\ref{kinetic.vs.potential}) by the energy balance between 
the two solutions. Only far below the Mott transition the superconductor is stabilized 
by a potential energy gain, as is the case in BCS theory. In the pseudogap regime,
corresponding to the expanded fullerides near the Mott transition, the stabilization 
is associated with a kinetic energy gain.

A similar phenomenon is well known in the optical conductivity of 
high-T$_c$ copper oxides~\cite{VanDerMarel}. Our calculations show that an 
increase in the zero-frequency optical conductivity or a kinetic 
energy gain do not actually exclude an electron-phonon pairing mechanism, 
but rather demonstrates the key importance of strong electronic correlations. 
Thanks to pairing, the motion of carriers in the superconducting phase is 
facilitated with respect to the pseudogap nonsuperconducting metal. In that
anomalous metal -- a non Fermi liquid-- the interaction constraints jam the 
free propagation of quasiparticles, causing a kinetic energy cost, partly released
with the onset of superconductivity. It would be of extreme interest 
if the optical conductivity increase demonstrated in cuprates could be investigated
in fullerides, both regular and expanded, since that would help 
discriminate between conventional BCS and SCS.

\begin{figure}[b]
\includegraphics[width=8cm]{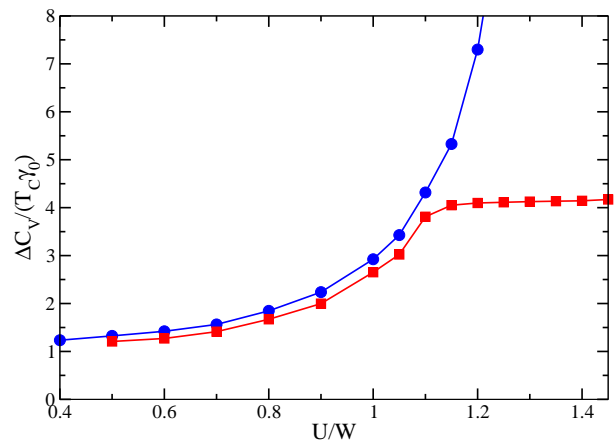}
\caption{\label{specific-heat-jump} Specific heat jump, in units of 
$T_c$, $\gamma_0$, as function of $U/W$. For comparison we also show the behavior 
of $1/Z(U)$, which should correspond to the same quantity if Fermi liquid theory were 
valid.}
\end{figure}

\section{Discussion of Experimental Probes}
\label{Sec:Comparison}
In the following we discuss whether and how the dominance of strong correlation 
we propose for the superconducting fullerides can be reconciled with the list of experimental facts 
given earlier, apparently supporting a conventional BCS behavior. The first quantity 
we discuss is the specific heat jump at $T_c$. Within BCS theory, the specific 
heat jump $\Delta C_V$ at $T_c$ is approximately 
\be
\frac{\displaystyle \Delta C_V }{\displaystyle T_c} \simeq 1.52\,\gamma_*,
\label{heat-jump}
\ee
where $\gamma_*$ is the specific heat coefficient of the metallic phase ($C_V = \gamma_* T$)), 
proportional to the mass enhancement $m^*/m$, in our case $1/Z$.  
The measured specific heat jump leads, through (\ref{heat-jump}), to an estimate 
of $\gamma_*\simeq 3\gamma_0$. This is indeed a rather low and regular value which has been 
commonly advocated as evidence of weak correlations in fullerenes. However, the standard 
argument only holds provided the normal phase is Fermi-liquid, which is not applicable 
close to a Mott transition, independently of any models. 

In Fig.~\ref{specific-heat-jump} we plot the jump in $C_V$ at $T_c$ for 
increasing $U/W$, compared with the Fermi-liquid estimate $1/Z$.\footnote{As mentioned in Sec. \ref{Sec:dmft}, 
for the present three-orbital model it has not been practical 
to include in Eq. (\ref{G_sum_Gm}) a number of excited states sufficient to make the truncation error 
in the Green's function negligible. 
For the data reported in Fig.~\ref{specific-heat-jump} we have an average error of 3-4\% 
which does not allow us to determine $T_c$ with sufficient accuracy. Yet, the jump of the specific heat relative to $T_c$ turns out to be almost 
independent on the truncation error and on the details of the calculations.}
After a region  
where the calculated quantity closely follows $1/Z$, $\Delta C_V/T_c$ flattens out 
and stays roughly constant around  $4\,\gamma_0$ up to the Mott 
transition, despite a diverging $1/Z$ (See Fig.~\ref{specific-heat-jump}). 
This shows that the energy scale that controls the superconducting instability 
is constant near the Mott transition, consistent with the single-impurity prediction 
that this scale should be identified by  $T_+$, a quantity of order $J$. 
The bottom line conclusion here is that a normally sized specific-heat jump does not imply
BCS superconductivity in fullerides.

\begin{figure}
\includegraphics[width=7cm]{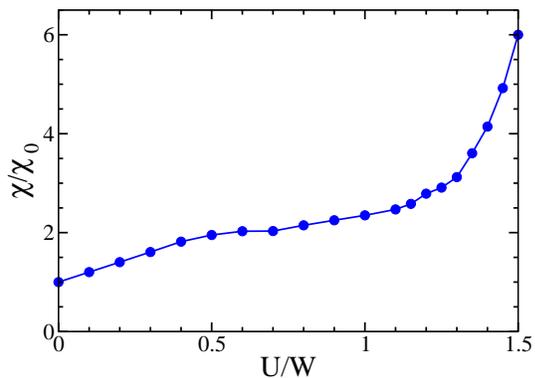}
\caption{\label{chi} Normal-phase uniform magnetic susceptibility $\chi$ normalized to the 
non-interacting value $\chi_0$. For $U=0$, $\chi\simeq \chi_0$, the difference being 
extremely small since $J\ll W$. The plateau between $U/W$ = 0.5 and 1 signals the 
effective crossover from a Fermi liquid with $S$= 3/2 per site, to a non Fermi liquid
with $S$=1/2 per site. }
\end{figure}

Another physical quantity which seemingly pointed towards weak 
correlations in fullerides is the magnetic susceptibility measured in the normal 
phase, apparently consistent with a weakly correlated a Fermi liquid and 
a Stoner enhancement of about a factor 2 to 3. 
This argument again becomes inconclusive once the Fermi-liquid 
scenario is abandoned. In Fig.~\ref{chi} we plot as function of $U/W$ the 
uniform magnetic susceptibility in the normal phase calculated by DMFT. 
After an initial Stoner enhancement at small $U/W$, the susceptibility flattens out 
and remains almost constant before a rapid growth which takes places extremely 
close to the Mott transition. In the plateau region the susceptibility 
enhancement of a factor between 2 and 3 with respect to $U=0$, surprisingly 
close to the experimental enhancement, covering the whole superconducting domain. 
Physically, the origin of this susceptibility plateau for increasing $U$
is quite instructive. It corresponds to the gradual crossover of the maximum 
spin available at each site from S=3/2 in the Fermi liquid at small $U < U_{cp}$
(Kondo screened AIM) to S=1/2 near Mott and large $U > U_{cp}$, where Fermi liquid 
behavior is lost. In essence, at large $U$ each molecule is effectively in a (dynamically) 
JT distorted state, where two out of three electrons are spin paired~\cite{Manini94}. 
We are thus led to conclude that the relatively weak observed enhancement of susceptibility 
does not correspond at all to a Stoner enhanced, weakly correlated Fermi liquid -- 
in fact quasiparticles do not even exist in most of the plateau region.
 
Finally, we wish to address signatures of the strongly correlated scenario which 
we expect should show up in important spectroscopies including 
the tunneling I-V characteristics and angle resolved photoemission spectroscopy (ARPES). 
This is initially embarrassing on two accounts. First, ARPES spectroscopies are $\mathbf{k}$-vector resolved, whereas
in DMFT we do not have access to any spatial structure. Second, tunneling spectroscopies 
are extremely well resolved near zero voltage, whereas our Lanczos method yields
a much poorer spectral function resolution in this region. 

Let us addressing tunneling first. Although Fig.~\ref{DOS} refers to the impurity 
C$_{60}^{3-}$ molecule, we believe that similar features would remain after full DMFT 
self-consistency in the normal phase -- if we only had a better low-frequency numerical resolution 
than we presently have. Hence, we suggest that tunneling I-V spectra of expanded 
fullerides be measured and examined, in order to bring out the expected rich structure 
of the kind sketched in Fig.~\ref{DOS}. 

Next, let us consider photoelectron spectroscopy. Again according to the single-impurity 
analysis~\cite{DeleoPRL}, the imaginary part of the single-particle self-energy should 
be finite and of order $T_+$ almost everywhere in the non-Fermi liquid normal phase above $T_c$. 
This has the following implications for ARPES:

\begin{itemize}

\item the fulleride photoemission spectrum should show $t_{1u}$ bands dispersing 
in the Brillouin zone with nonzero bandwidth, governed by the energy scale $T_+$. 
The value of $T_+$ decreases with increasing $U/W$ (increasing expansion), from 
$W$ at $U$ = 0 to (larger than) $J$ at the Mott transition; 

\item there should be a spectral peak broadening of the same order 
of magnitude $T_+$ as the apparent $\mathbf{k}$-resolved band dispersion. In particular the broadening 
should remain constant approaching the Fermi surface -- unlike a Fermi liquid phase 
where quasi-particle peaks become narrower and narrower.

\end{itemize}
\begin{figure}
\includegraphics[width=7cm]{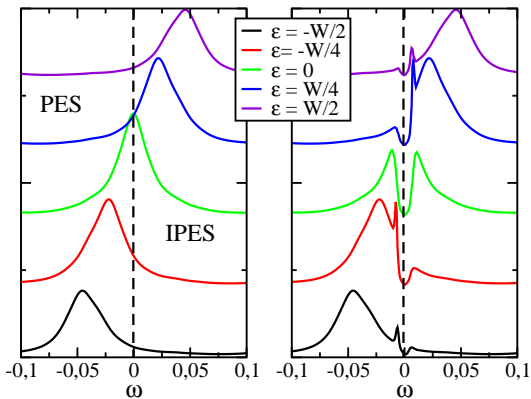}
\caption{\label{Arpes} Simulated photoemission spectra elaborated form the 
DMFT result in the normal phase. The left panel refers to $U/W = 1.1$ (corresponding 
to unexpanded fullerides), while the right panel is for $U/W = 1.3$ (corresponding to 
expanded fullerides).
}
\end{figure} 

The momentum-independence of the DMFT self-energy implies that, in this approximation, the 
$\mathbf{k}$-modulation of the electronic dispersion is assumed to remain unaffected by interactions. 
Within this assumption and without requiring too high frequency-accuracy, we can compute a toy $\mathbf{k}$-resolved spectral function according to 

\begin{equation}
\label{eq:arpes}
A(k,\omega)= -\frac{1}{\pi}  
\mbox{Im} \frac{1}{\omega-\varepsilon_{\bf k} -\Sigma_{DMFT}(\omega)},
\end{equation}

where $\varepsilon_{\bf k}$ is the non interacting dispersion and $\Sigma_{DMFT}(\omega)$ is the DMFT self-energy calculated with a finite 
number of baths. The effect 
of the local self-energy will be to change the effective bandwidth and to give 
rise to finite lifetime effects, even if the $\mathbf{k}$-modulation of the dispersion 
is unrenormalized. For our Bethe lattice, there is no straightforward definition 
of momentum, and we remedy that by computing $A(\varepsilon,\omega)$, 
which corresponds to Eq. (\ref{eq:arpes}) with $\varepsilon_{\bf k} \to \varepsilon$.

In Fig.~\ref{Arpes} we show theoretical ARPES results for some choices of 
$\varepsilon$ obtained by using the DMFT self-energy for temperatures above 
$T_c$, both for a value of $U/W =1.1$ which lies close to the maximum of the 
superconducting dome, but still on the less correlated side, corresponding 
to unexpanded (or moderately expanded) fullerides, and for a value which 
lies in the downward branch of the dome ($U/W = 1.3$), corresponding to a
very expanded fulleride. We note in both cases the existence of an incoherent 
low-energy feature dispersing with a reduced but nonzero electron bandwidth 
of $0.1~W$. In the expanded case the pseudogap feature is clearly present.

Recent photoemission spectra of K$_3$C$_{60}$~\cite{Goldoni} indicate an
overall dispersion bandwidth of about 160~meV, about a quarter of the bare 
calculated bandwidth in the local-density approximation. The experimental spectral 
peak does not show the usual Fermi-liquid-like narrowing on approaching the Fermi 
level, a fact which is in agreement with our expectation for a non Fermi liquid 
(although a non expanded fulleride like K$_3$C$_{60}$ probably lies at the 
beginning of the SCS dome, where deviations from Fermi liquid are not massive). 
Experimentally the peak does not appear to cross the Fermi level, and the 
intensity instead drops, suggestive of a pseudogap. Unfortunately the spectrum
shows very strong vibronic effects, reflecting the retarded strong electron Jahn Teller 
coupling. This aspect is not covered by our unretarded approximation, but it
heavily affects the line shape and hampers the extraction of purely electronic features. 
Treatment of the vibronic effects, and a quantitative description
of dispersion, will require abandoning in the future our approximation of infinitely 
fast phonon dynamics, as well as a possible extension to cluster extensions of DMFT which 
allow for different renormalizations of different momenta~\cite{cdmft}.

\section{Conclusions}
Summarizing, we addressed the apparently contradictory properties
of expanded trivalent fullerides superconductors and insulators -- and to some extent 
of the whole family of fullerides -- and presented 
a theoretical scenario emphasizing the role of strong electron correlations. 
That is especially designed and appropriate for the more expanded
members of the family, such as (NH$_3$)$_x$NaK$_2$C$_{60}$, Li$_3$C$_{60}$, 
Cs$_{3-x}$K$_x$C$_{60}$ and Cs$_{3-x}$Rb$_x$C$_{60}$, and the recently discovered A15 $Cs_{3}C_{60}$ 
that are near or past the Mott transition.

Our model explains the dome-shaped increase and subsequent decrease of $T_c$ 
upon expansion of the lattice spacing in fullerides; the coexistence 
of metallic behavior and of Mott insulator features such as the large NMR spin 
gap in all fullerides, and the $S=1/2$ spin in the insulating state (identified 
as a Mott-Jahn-Teller insulator). It explains why the $s$-wave $T_c$ can be 
as high as 40~K even though the Coulomb interaction strength is prohibitive, and
why $T_c$ does not automatically decrease upon increase of $U/W$. It also accounts 
for more standard observations, such as regular specific heat jumps and moderately 
high spin susceptibilities, facts that were so far construed as evidence for conventional 
BCS superconductivity. 

Besides those listed in the previous Section, one can anticipate a number of additional experiments 
that could provide ``smoking gun'' evidence for strongly correlated superconductivity
in fullerides. The tunneling $I$-$V$ characteristics observable, e.g., by a scanning tunneling spectroscopy tip 
should, in an expanded fulleride, develop the low energy
features typical of the Kondo impurity spectral function. The isotope effect
upon carbon substitution should also behave very unconventionally, and eventually get 
smaller as the superconducting dome is passed and the Mott transition is approached 
upon expansion. In this regime, as the quasiparticle bandwidth $ZW$ gradually falls below
the typical energy $\hbar\omega$ of an increasing fraction of the eight $H_g$ 
Jahn Teller modes, the associated retardation effect should in fact disappear.
%
%
The expanded fullerides and related materials, clearly not enough
investigated so far, deserve in our view the strongest experimental attention.
They combine elements that make them members at large of the high temperature 
superconductor family. They combine neighborhood of the Mott transition and predominance 
of strong electron correlations, with conventional elements such as electron-phonon $s$-wave 
pairing, that are typical of BCS systems. Our study identifies a pseudogap and other 
features in the IV tunneling spectrum, an increase of zero-frequency optical weight in the optical 
response of the superconducting phase, and the emergence of two separate energy scales 
in ARPES as the most urgent experimental undertakings that could confirm 
of falsify our claims.

\section*{Acknowledgments}
E.T. thanks Kosmas Prassides, Andrea Goldoni, and Mauro Ricco' for information 
and discussion about fullerides and again K.P. for providing us with Fig. 1. 
M.C. acknowledges discussions with A. Toschi. Work in SISSA was sponsored by PRIN 
Cofin 2006022847, as well as by INFM/CNR ``Iniziativa trasversale calcolo parallelo''. 
Work in Rome was sponsored by PRIN Cofin 200522492.


\end{document}